# Attempt to Predict Failure Case Classification in a Failure Database by using Neural Network Models

Koichi Bando[1], Kenji Tanaka[1]
[1] Graduate School of Informatics and Engineering, The University of Electro-Communications, Tokyo, Japan
1-5-1 Chofugaoka, Chofu-Shi, Tokyo 182-8585, Japan
E-mail: k-bando@mth.biglobe.ne.jp, tanaka@is.uec.ac.jp

*Abstract*— With the recent progress of information technology, the use of networked information systems has rapidly expanded. Electronic commerce and electronic payments between banks and companies, and online shopping and social networking services used by the general public are examples of such systems. Therefore, in order to maintain and improve the dependability of these systems, we are constructing a failure database from past failure cases. When importing new failure cases to the database, it is necessary to classify these cases according to failure type. The problems are the accuracy and efficiency of the classification. Especially when working with multiple individuals, unification of classification is required. Therefore, we are attempting to automate classification using machine learning. As evaluation models, we selected the multilayer perceptron (MLP), the convolutional neural network (CNN), and the recurrent neural network (RNN), which are models that use neural networks. As a result, the optimal model in terms of accuracy is first the MLP followed by the CNN, and the processing time of the classification is practical.

## I. INTRODUCTION

With the progress of information technology (IT), the use of networked information systems has rapidly expanded. Examples of such systems include electronic commerce and electronic payments between banks and companies and the use of online shopping and social networking services (SNS) by the general public. Therefore, in order to maintain and improve the dependability of information systems, the authors are constructing a failure database (DB) by collecting numerous past failure cases that can be referred to by system developers and operators in order to prevent the recurrence of failures and prevent new failures [1, 2]. When importing new failure cases, it is necessary to classify the cases according to failure type. At present, the classification is performed by individuals. The problems are the accuracy and efficiency of the classification. Especially when working with multiple people, unification of the classification is required. In order to address these problems, the authors are attempting to automate the classification by machine learning using a neural network model. The following describes previous research related to machine learning using neural networks.

Kim [3] reported experiments with a convolutional neural network (CNN) built on top of Word2Vec for classification tasks and showed the results on benchmarks. Goh et al. [4] applied the multilayer perceptron (MLP) to web spam classification and showed that the MLP outperforms the support vector machine. Dhingra et al. [5] reported a spam classification approach using MLP learning, and the results were compared with previous techniques.

We selected the MLP, the CNN, and the recurrent neural network (RNN) as the neural network model, and compared and evaluated the accuracy of the resulting classifications in order to clarify the optimum model in terms of accuracy. Our contribution is to improve the accuracy and efficiency of the classification by automating the classification. Furthermore, we confirm the practicality in terms of processing time.

## II. SYSTEM CONSTRUCTION AND TOOL DEVELOPMENT

The failure DB used in the present paper was extracted from the failure DB accumulated in our previous research, focusing on communication network failures and financial information system failures from 2000 to 2015 in Japan [1, 2]. The sources of the failure DB were publicly available information such as newspapers, journals. The extracted failure data are classified according to the phenomenon. Though Avizienis et al. [6] provided a detailed classification of the faults, our classification is made at the failure level to classify failures in terms of phenomena from the user's point of view. When performing cause analysis, we used the classification of faults by Avizienis et al. Table 1 shows our classification of failures that consists of major classes that can be used in common regardless of the field and subclasses based on typical failure cases in each field.

As a development tool, we introduced "Janome" [7] as a morphological analysis library, and Tensorflow [8] and Keras [9] as deep learning environments. As models that use a neural network for learning and predicting classification, we use MLP, CNN and RNN. The MLP is composed of an input layer, hidden layers (2 dropouts, a rectified linear unit), and an output layer. After text processing, such as word-separation in Japanese, the text is input into the input layer. The Output layer outputs accuracy as a learning result. The predicted value for each test data is output by executing using the learning result. The CNN is composed of an embedding layer, hidden layers (a convolutional, a pooling, a rectified linear unit, etc.)

Table 1 Summary of failure data

| Fields | Classifications of types of failures | | ID | No. of failures | No. of test cases*1 |
|---|---|---|---|---|---|
| | Major class | Subclass | | | |
| Communication | service-related | telecom service suspended | C-A1 | 510 | 41 |
| | | telecom service quality impaired | C-A2 | 122 | 10 |
| | | partial malfunction | C-A3 | 214 | 17 |
| | processing-related | misclaim of charges | C-B1 | 78 | 6 |
| | information-related | information leakage (mistakes) | C-C1 | 71 | 6 |
| | | data loss, incorrect registration | C-C2 | 7 | 0 |
| | equipment-related | malfunction | C-D1 | 114 | 9 |
| | | safety problem | C-D2 | 33 | 4 |
| | cybercrime-related | information leakage (crime) | C-E1 | 50 | 4 |
| | | information security crimes | C-E2 | 30 | 3 |
| | other | other | C-F1 | 9 | 0 |
| | | Total | | 1238 | 100 |
| Finance | service-related | all service stoppage | F-A1 | 38 | 3 |
| | | terminal stoppage | F-A2 | 193 | 14 |
| | | partial malfunction | F-A3 | 231 | 17 |
| | | Internal services not available | F-A4 | 213 | 16 |
| | cybercrime-related | information leakage (crime) | F-E2 | 37 | 3 |
| | | information security crimes | F-E3 | 24 | 2 |
| | other | other | F-F1 | 0 | 0 |
| | | Total | | 1347 | 100 |
| | | Grand total | | 2585 | 200 |

*1 The number of cases in this column is the number in the left column.

and an output layer. The input to the embedding layer is a word vector generated by word embedding using the Word2Vec model [10]. The RNN is composed of an embedding layer as a CNN, hidden layers (a long short-term memory, rectified linear unit, etc.), and an output layer.

III. IMPLEMENTATION RESULTS

Fig. 1 shows the accuracy of the prediction results of the classifications for the major class and subclass of each model. For the evaluations, the test cases in Table 1 were used. These results are the average values implemented five times. As a result, the major class accuracy is 97.4% for the CNN, 96.9% for the MLP, and 95.3% for the RNN. Here, the accuracy of the CNN is the best, but the difference in accuracy from the second-ranked MLP is very small. The subclass accuracy is 92.4% for the MLP, 89.8% for the CNN, and 84.4% for the RNN. Here, the accuracy of the MLP is the best. The accuracy of the MLP is notably high, even for data in which multiple fields are mixed, such as communication and finance. As for the processing time, the common processing time in total, such as the text processing time and the learning time, was 794 s, and the individual processing time was 0.5 s per inquiry. The PC model used was a DELL Precision 7920 Tower. These results demonstrate that the processing time is practical.

IV. DISCUSSION

In order to clarify the cause of the difference in accuracy for each of the above models, the number of mismatch cases with the correct answer in the implementation results is aggregated for the field, major class, and subclass, and the mismatch rate is calculated (Fig. 2). In the present paper, the fields of interest are communications and finance. From Fig. 2, the field mismatch rates due to this factor are the CNN/RNN and the MLP in ascending order, and the differences are small. More than half of these are incorrect cases in the same major class in different fields. The major class mismatch rates by model are the CNN, the MLP, and the RNN in ascending order. The cause of these mismatches is mistakes in high-similarity cases. The subclass mismatch rates by model are the MLP, the CNN and the RNN in ascending order. The CNN and RNN have higher mismatch rate than the MLP. The cause of these mismatches is mistakes in high-similarity cases.

The difference in the mismatch rate of each model is considered to be caused by the features of the input. The input of the MLP is a count-based method based on the frequency of occurrence of words and statistical processing of the entire DB performed by TF-IDF. On the other hand, the CNN and the RNN are based on the word vector generated by Word2Vec with a similar meaning. Therefore, regarding subclass mismatches, it is considered that the mismatch rate is larger than the count-based MLP because it is difficult to make a difference in the classification prediction of cases that are semantically similar to each other, such as a set of cases in the same major class. This difficulty in prediction is thought to be related to the text structure in the DB. In the present paper, the text of each failure case consists of simple sentences that describe the facts. Therefore, it is considered that the advantages of the word vectors used in the CNN and the RNN, such as the reflection of word order, are not fully utilized.

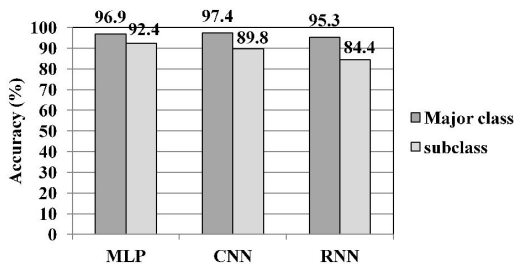

Fig.1 Accuracy of major class and subclass predictions for each model

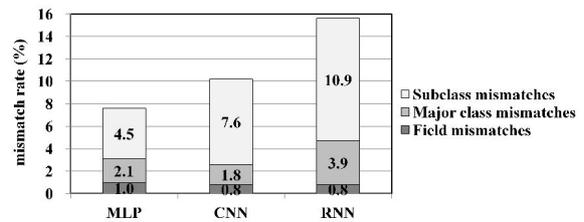

Fig.2 Mismatch rate for each model

V. CONCLUSION

First, we conducted learning on each model using the classification information of the failure DB as training data and evaluated the accuracy of the classification prediction using 200 test cases. The subclass accuracy was 92.4% for the MLP, 89.8% for the CNN, and 84.4% for the RNN. In order to clarify the factors involved in the differences among models, we analyzed the mismatched cases for each model in detail. As mentioned in Section IV, the major factor was determined to be the differences in input features. Based on the above considerations, the optimal model for the classification prediction regarding the failure DB is the MLP, followed by the CNN. Areas for future study include the examination of the optimal model for different types of DB.

ACKNOWLEDGEMENT

The present study was supported by JSPS through a Grant in-Aid (19H02386).